\documentclass[11pt]{article}

\usepackage[utf8]{inputenc}
\usepackage[T1]{fontenc}
\usepackage{amsmath,amssymb,amsthm}
\usepackage{graphicx}
\usepackage{hyperref}
\usepackage{microtype}
\usepackage[margin=1in]{geometry}
\usepackage{newtxtext,newtxmath}
\usepackage{booktabs}

\graphicspath{{figures/}}

\theoremstyle{plain}

\theoremstyle{definition}

\theoremstyle{remark}

\title{Infinite Canons: Maximally Self-Similar Melodic Lines and Canons with Infinite Solutions}
\author{Clifton Callender\thanks{Earlier versions of this material were presented as ``\emph{Spira mirabilis}, for player piano,'' Joint Mathematics Meetings, New Orleans, January 6, 2011 (invited presentation; abstract 1067-00-1969, \emph{Abstracts of Papers Presented to the American Mathematical Society} 32, no.~1); ``Infinitely variable tempo canons,'' Nancarrow in the
21st Century International Conference, Southbank Centre / Trinity Laban Conservatoire of Music and Dance, London, April 21--22, 2012; and ``Maximally Self-Similar Melodies and Canons with Infinite Solutions,'' joint annual meeting of the American Musicological Society, Society for Ethnomusicology, and Society for Music Theory, New Orleans, November 1--4, 2012. Related material is posted at \url{https://cliftoncallender.com/research/infinite_canons/}. An expanded and more developed treatment of these topics is in preparation.}\\
\small Florida State University}
\date{\today}

\begin{document}
\maketitle

\begin{abstract}
\textit{Infinite Canons} is an ongoing series of canons with infinite solutions. More specifically, each canon is based on a melodic line that can be combined in any number of voices, at any tempo ratios (rational or irrational), and with each voice moving either forward or in retrograde inversion, while maintaining harmonic consistency. This paper describes the structure of these maximally self-similar melodic lines based on two different constructions: 1) a discrete prime-factorization approach yielding self-similarity under all rational ratios, and 2) a continuous logarithmic approach extending this to irrational ratios. In both cases the vertical interval between voices in a tempo ratio of $\lambda_i / \lambda_j$ is given by a homomorphism $\phi(\lambda_i / \lambda_j)$, which is a constant independent of time. Furthermore, under these constructions retrograde inversion collapses to transposition, allowing for all manner of table canons. These structures are demonstrated with suggestive realizations of several different infinite canons. Future work includes a more complete mathematical treatment, musical applications, and an interactive program that allows users to explore an unlimited number of realizations of these pieces.
\end{abstract}

% =====================================================================

\section{Introduction}
\label{sec:intro}

\begin{figure}[htbp]
  \centering
  \includegraphics[width=\textwidth]{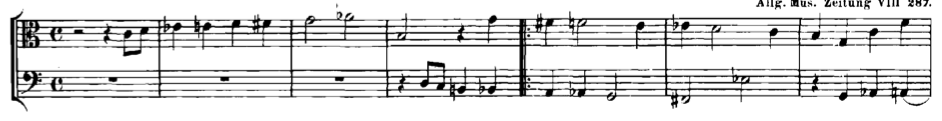}
  \includegraphics[width=\textwidth]{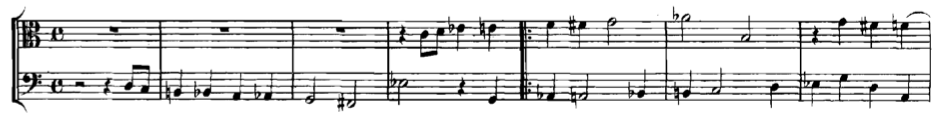}
  \caption{Two solutions for the ``Quaerendo invenietis'' canon from J.~S. Bach's \emph{Musical Offering}.}
  \label{fig:bach}
\end{figure}

A musical canon is a composition in which a melodic line is placed in counterpoint with itself according to a rule combining temporal shift, transposition, inversion, retrograde, and/or augmentation or diminution. Most canons have a single solution---a single rule by which the line combines with itself successfully---but some canons have more. Figure~\ref{fig:bach} shows
the two non-trivially distinct solutions for the two-voice canon marked ``Quaerendo invenietis'' from J.~S. Bach's \emph{Musical Offering}. The existence of canons that work under multiple rules raises an interesting question: Is there a limit to the number of solutions a canonic line can possess?

One of Conlon Nancarrow's more curious works is Study No.~44, which Kyle Gann~\cite{gann} calls the ``Aleatory Canon.'' The canon is for two voices in any tempo ratio. (Each voice of the canon is played on a different piano using a separate roll.)
Nancarrow maintains \emph{global} harmonic control by keeping his materials very simple. (The two voices combined mostly stay within a single diatonic collection.) However, \emph{local} harmonic details are essentially unplanned, since the exact alignment of the two canonic voices is unknowable ahead of time. This raises another question: Is it possible to create a canon that works in any tempo ratio while maintaining control of local-level harmonic details?

This paper provides an answer to both questions by constructing melodic lines that can be combined in any number of voices, in any tempo ratios (rational or irrational), and with each voice moving either forward, or backward and upside down through the line, while maintaining harmonic consistency. The key is that the canonic line must be maximally self similar; that is, the melody must be self similar at \emph{all possible time scales}.

\begin{figure}[htbp]
  \centering
  \includegraphics[width=0.8\textwidth]{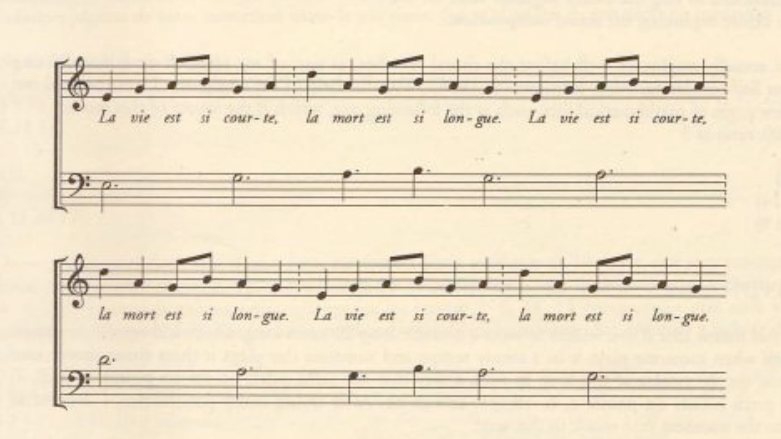}
  \caption{Tom Johnson's \emph{La vie est si courte} is self similar at the ratio 3:1.}
  \label{fig:johnson}
\end{figure}

\begin{figure}[htbp]
  \centering
  \includegraphics[width=0.8\textwidth]{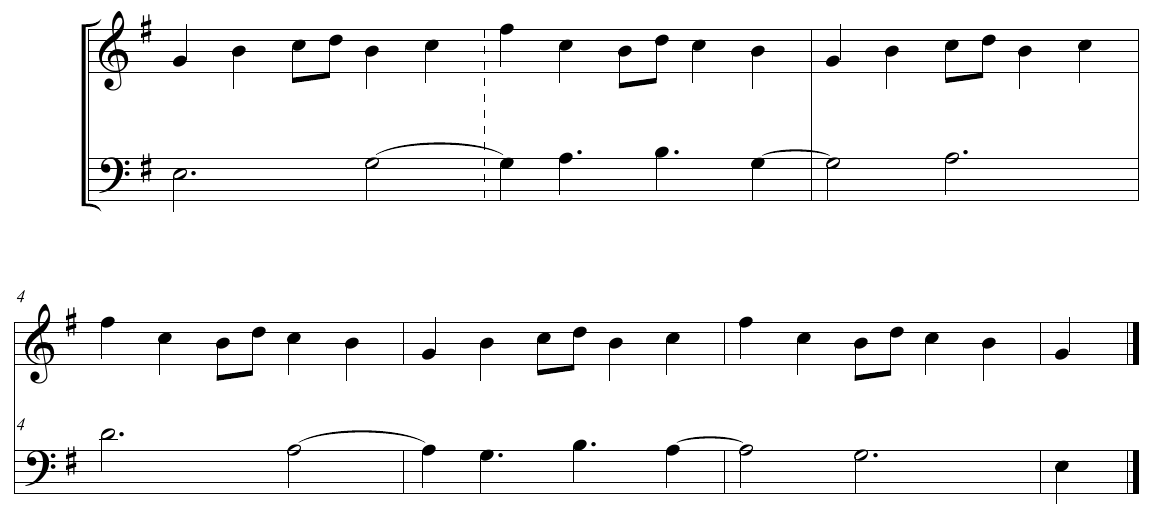}
  \caption{Tom Johnson's \emph{La vie est si courte}. Here the self-similarity is manifest as a 3:1 tempo canon where simultaneous attacks form a series of parallel tenths.}
  \label{fig:johnson-tenths}
\end{figure}

Tom Johnson's \emph{La vie est si courte}, shown in Figure~\ref{fig:johnson}, is an example of self similarity at a single time scale. The lower line is the same as the upper line played three times as slow. Note that the beginning of every note in the lower line forms an octave with the upper line. So the melody of \emph{La vie est si courte} is self similar at the ratio of 3:1. Equivalently, the melody gives rise to a 3:1 tempo canon that exhibits maximal harmonic consistency---a series of parallel octaves. We can enhance the independence of the two lines while obscuring the melody's self-similar structure by transposing one of the voices. For instance, transposing the upper line of this 3:1 tempo canon up a diatonic third (assuming an implicit key signature of one sharp) yields a series of parallel tenths, shown in Figure~\ref{fig:johnson-tenths}. Johnson~\cite{johnson}, Emmanuel Amiot~\cite{amiot}, and others have written about these self-similar melodies at finite time scales (see \S\ref{sec:2.7}), but we will need to use a different approach to extend this important idea to all possible time scales, with the closest prior treatment being Hendriks et al.~\cite{hendriks2012}, discussed in \S\ref{sec:2.4}.

This paper details the construction of maximally self-similar lines in two cases. Section~\ref{sec:discrete} constructs such lines in the discrete case, allowing for infinite combinations at all rational tempo ratios, by freely assigning directed intervals to every prime number and using prime factorization to define the map from time points to directed intervals $ \phi \colon (\mathbb{Q}^{+}, \times) \longrightarrow (\mathbb{R}, +)$. Intervals are determined entirely by the ratios of time points, which is the necessary and sufficient condition for maximal self similarity. Section~\ref{sec:continuous} extends this approach through Cauchy's logarithmic functional equation to the continuous case, including irrational ratios. Throughout both Sections~\ref{sec:discrete} and~\ref{sec:continuous}, we consider structural properties of these lines, including issues of periodicity and aperiodicity and the temporal relationship between time points and harmonic rhythm.

The musical significance of these structures is demonstrated in Section~\ref{sec:examples} by two canons based on the same lines with tempo ratios $4\!:\!2\!:\!1$ and $\pi\!:\!e\!:\!1$, respectively. Maximally self-similar lines naturally give rise to canons by retrograde inversion, sometimes called table canons, which is the subject of Section~\ref{sec:table}, revealing a strong connection between retrograde and inversion. A concluding section points to future elaborations, including a greater variety of musical realizations, recent work by Dmitri Tymoczko, and more thorough engagement with the underlying mathematics, including the work of Hendriks et al.\ and proof of the extension of their aperiodicity results.

% =====================================================================

\section{The Discrete Construction}
\label{sec:discrete}

\subsection{}
\label{sec:2.1}

\begin{figure}[htbp]
  \centering
  \includegraphics[width=\textwidth]{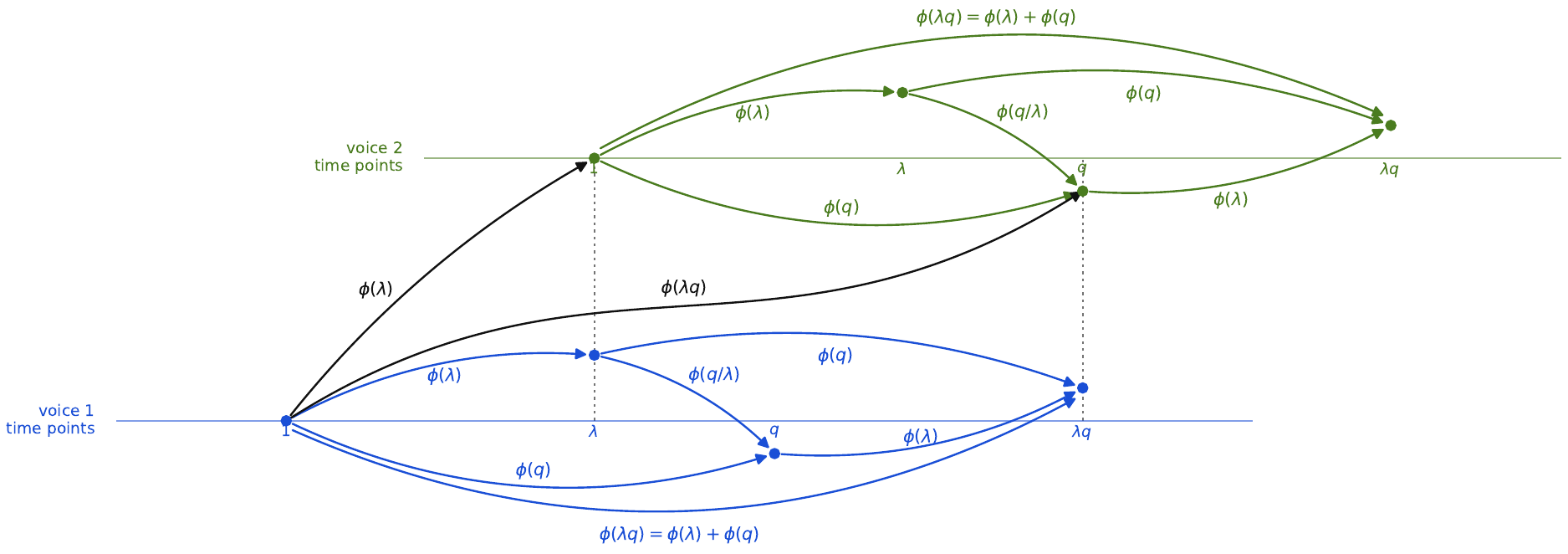}
  \caption{Diagram of a two-voice four-note canon by augmentation at the rational factor $\lambda$. In order for simultaneous time points to form a unison (and thus ensure self-similarity), the second voice (green) must be a transposition of the first (blue) by the interval $\phi(\lambda)$. Intervals relate to time points and combine according to the homomorphism $\phi(\lambda q) = \phi(\lambda) + \phi(q)$.}
  \label{fig:homomorphism}
\end{figure}

In this section we detail a discrete approach to constructing melodic lines that are self similar at all positive rational time scales. These are melodies that can combine with any rational augmentation or diminution to create a tempo canon maintaining harmonic consistency. Consider the two-voice four-note canon diagrammed in Figure~\ref{fig:homomorphism}, where the second voice is an augmentation of the first by a factor of $\lambda$. This means that every time point, $t$, in the second voice coincides with the first at $\lambda t$. In particular, time points $1$ and $q$ in the second voice coincide with time points $\lambda$ and $\lambda q$ in the first. We indicate the pitch interval from time point $1$ to $\lambda$ in the first voice by $\phi(\lambda)$, using the ratio of time points, $\lambda / 1 = \lambda$. In order for time point $1$ in the second voice to form a unison with the first, the entirety of the augmented voice must be transposed by this same interval, $\phi(\lambda)$. Note also that the interval from $1$ to $\lambda q$ in voice one (and to $q$ in voice two), $\phi(\lambda q)$, is equal to the sum of the interval for each factor: $\phi(\lambda q) = \phi(\lambda) + \phi(q)$. This is the essential homomorphism from the multiplicative realm of positive, rational time points and the additive realm of real-valued intervals, and the crux of our construction:
\begin{equation}
  \phi \colon (\mathbb{Q}^{+}, \times) \longrightarrow (\mathbb{R}, +).
  \label{eq:hom}
\end{equation}
Since each rational time point, $q \in \mathbb{Q}^{+}$, has a unique prime factorization, we can fully determine a maximally self-similar melody by defining a real-valued interval for each prime number.\footnote{Hendriks et al.~\cite{hendriks2012} study one-sided infinite sequences over a finite alphabet. Restricted to integer time points and to pitch classes rather than pitches, our sequences are identical to theirs. See \S\ref{sec:2.4} for more on the relationship with their work.}

\subsection{}
\label{sec:2.2}

Let $\tau(p) \in \mathbb{R}$ be the freely chosen directed interval associated with prime $p$. For any time point
$q = \prod_{p} p^{\,a_p}$, $a_p \in \mathbb{Z}$, the interval from time point $1$ to $q$ is
\begin{equation}
  \phi(q) \;=\; \sum_{p} a_p\, \tau(p).
  \label{eq:phi}
\end{equation}
(In particular $\phi(p) = \tau(p)$ on primes.) Assuming a reference pitch $s$ at time point $1$, where $\phi(1) = 0$, the pitch at time point $q$ is
\begin{equation}
  T_{\phi(q)}(s).
  \label{eq:pitch}
\end{equation}

This completes the construction of a maximally self-similar melodic line. As discussed in \S\ref{sec:2.1}, to maintain harmonic unisons at coincident time points in augmentation canon by $\lambda$, the augmented voice must be transposed by $T_{\phi(\lambda)}$. More generally, for an $n$-voice tempo canon with voices in the ratios $\lambda_1, \lambda_2, \ldots, \lambda_n$, where $\lambda_i$ is the particular ratio for voice $i$, the interval from voice $i$ to voice $j$ at a global time point $t$ is given by
\begin{equation}
  \phi\!\left(\frac{t}{\lambda_j}\right)
  - \phi\!\left(\frac{t}{\lambda_i}\right)
  \;=\; \phi\!\left(\frac{\lambda_j}{\lambda_i}\right).
  \label{eq:constant-interval}
\end{equation}
That is, the harmonic interval between voices is a constant independent of time, which is precisely the lever for exerting harmonic control between voices in all possible rational tempo ratios. Since the harmonic interval is a constant, we can transpose the voices relative to one another to achieve the desired harmonic combinations.

\subsection{}
\label{sec:2.3}

\begin{figure}[htbp]
  \centering
  \includegraphics[width=\textwidth]{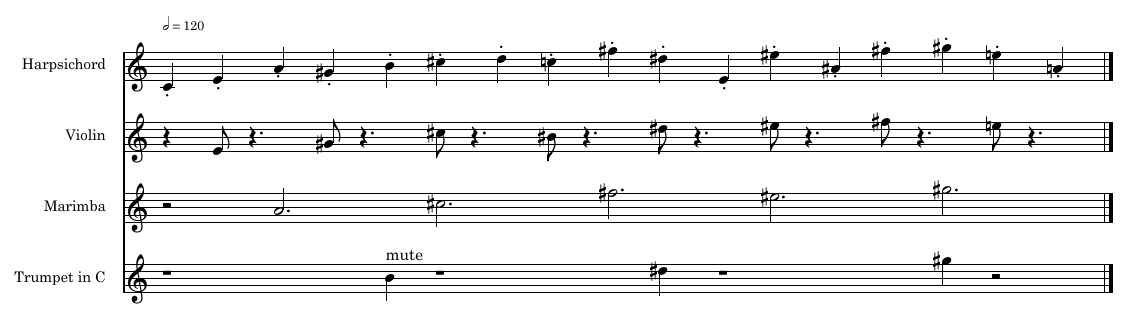}
  \caption{Self-similarity canon by ratios $1\!:\!2\!:\!3\!:\!5$.}
  \label{ex:one}
\end{figure}

The melody in Example~\ref{ex:one} serves to demonstrate the formalism through a tangible musical application. The melody is stated in full in the top staff (harpsichord) and is based on the following mappings of primes to directed intervals (measured in semitones): $\tau(2) = 4$,  $\tau(3) = 9$,  $\tau(5) = 11$, $\tau(7) = 14$, $\tau(11)= 4$,  $\tau(13) = 10$, and $\tau(17) = 9$. This mapping combined with the reference pitch of $C_4$ at time point $1$ fully determines the pitches for every time point in the melody, as shown in Table~\ref{tab:intervals}.
\begin{table}[htbp]
  \centering
  \begin{tabular}{@{}ccl@{}}
    \toprule
    Time point & Factors & Interval from time point 1 \\
    \midrule
     1 & $2^{0}3^{0}5^{0}\cdots$ & $0$ \\
     2 & $2^{1}$          & $4$ \\
     3 & $3^{1}$          & $9$ \\
     4 & $2^{2}$          & $4 + 4 = 8$ \\
     5 & $5^{1}$          & $11$ \\
     6 & $2 \cdot 3$      & $4 + 9 = 13$ \\
     7 & $7^{1}$          & $14$ \\
     8 & $2^{3}$          & $4 + 4 + 4 = 12$ \\
     9 & $3^{2}$          & $9 + 9 = 18$ \\
    10 & $2 \cdot 5$      & $4 + 11 = 15$ \\
    11 & $11^{1}$         & $4$ \\
    12 & $2^{2} \cdot 3$  & $4 + 4 + 9 = 17$ \\
    13 & $13^{1}$         & $10$ \\
    14 & $2 \cdot 7$      & $4 + 14 = 18$ \\
    15 & $3 \cdot 5$      & $9 + 11 = 20$ \\
    16 & $2^{4}$          & $4 + 4 + 4 + 4 = 16$ \\
    17 & $17^{1}$         & $9$ \\
    \bottomrule
  \end{tabular}
  \caption{Time points, prime factorization, and directed intervals for the melody constructed in Example~\ref{ex:one}.}
  \label{tab:intervals}
\end{table}
Note that the violin, marimba, and trumpet are augmentations of the
harpsichord by factors of $2$, $3$, and $5$, respectively, and that
these lines are transpositions of the melody by the respective intervals
$\phi(2)$, $\phi(3)$, and $\phi(5)$. This construction ensures that all
simultaneous attacks form a unison, manifesting the self-similar
property at multiple time scales.

\subsection{}
\label{sec:2.4}

Hendriks et al.~\cite{hendriks2012} study one-sided infinite sequences over a finite alphabet, proving that arithmetically self-similar sequences (up to a ``vertical shift'') are completely additive, where
$f(mn) = f(m) + f(n)$. Their scale-invariance, where $w(kn) - w(n)$ is a constant, with $k \in \mathbb{Z}$ and $n \in \mathbb{N}^{+}$, is the same as the finding that directed intervals in our construction are independent of time, $\phi(\lambda q) - \phi(q) = \phi(\lambda)$. Their vertical shift, which is addition by a constant in a finite cyclic group, relates to tempo canons, where augmentation by $\lambda$ corresponds to a transposition by $\phi(\lambda)$.

The current paper extends Hendriks et al.\ in several respects. First, while the domain for their sequences is the positive naturals ($\mathbb{N}^{+}$), our domain is all positive rationals ($\mathbb{Q}^{+}$). This is what allows our time points to include rational subdivisions of integer downbeats. It also allows for time points and augmentations less than one, where $\lambda < 1$ would be perceived as a diminution. Time point zero is a rest, since there are no combinations of prime factors that yield $0$. Second, while their codomain is restricted to a finite cyclic group ($\mathbb{Z}/n\mathbb{Z}$), our codomain is $\mathbb{R}$, which means that the constructed melodies are self similar with respect to pitch, not just pitch class, and are not restricted to a fixed equal division of the octave.

\subsection{}
\label{sec:2.5}

In the introduction, we emphasized that these canons have an infinite number of solutions, but they are also potentially infinite as aperiodic sequences of arbitrary length. Hendriks et al.\ prove that the only completely additive (and therefore self-similar) sequence satisfying $w(n + P) = w(n)$ for some period $P$ is the trivial sequence of constant value $w(n) = 0$ for all $n$. Musically, this means that the only periodic, maximally self-similar line is one that sounds a constant pitch for every time point. This finding can be strengthened (proof deferred) to exclude sequences in which a regularly recurring block of length $P$ repeats with an arbitrary transposition for each block. Again, any sequence that satisfies this more relaxed condition,
\begin{equation}
  \phi(n + P) \;=\; \phi(n) + C_{\lfloor n/P \rfloor},
  \label{eq:relaxed-period}
\end{equation}
must be a trivial sequence of constant value.

\subsection{}
\label{sec:2.6}

\begin{figure}[htbp]
  \centering
  \includegraphics[width=\textwidth]{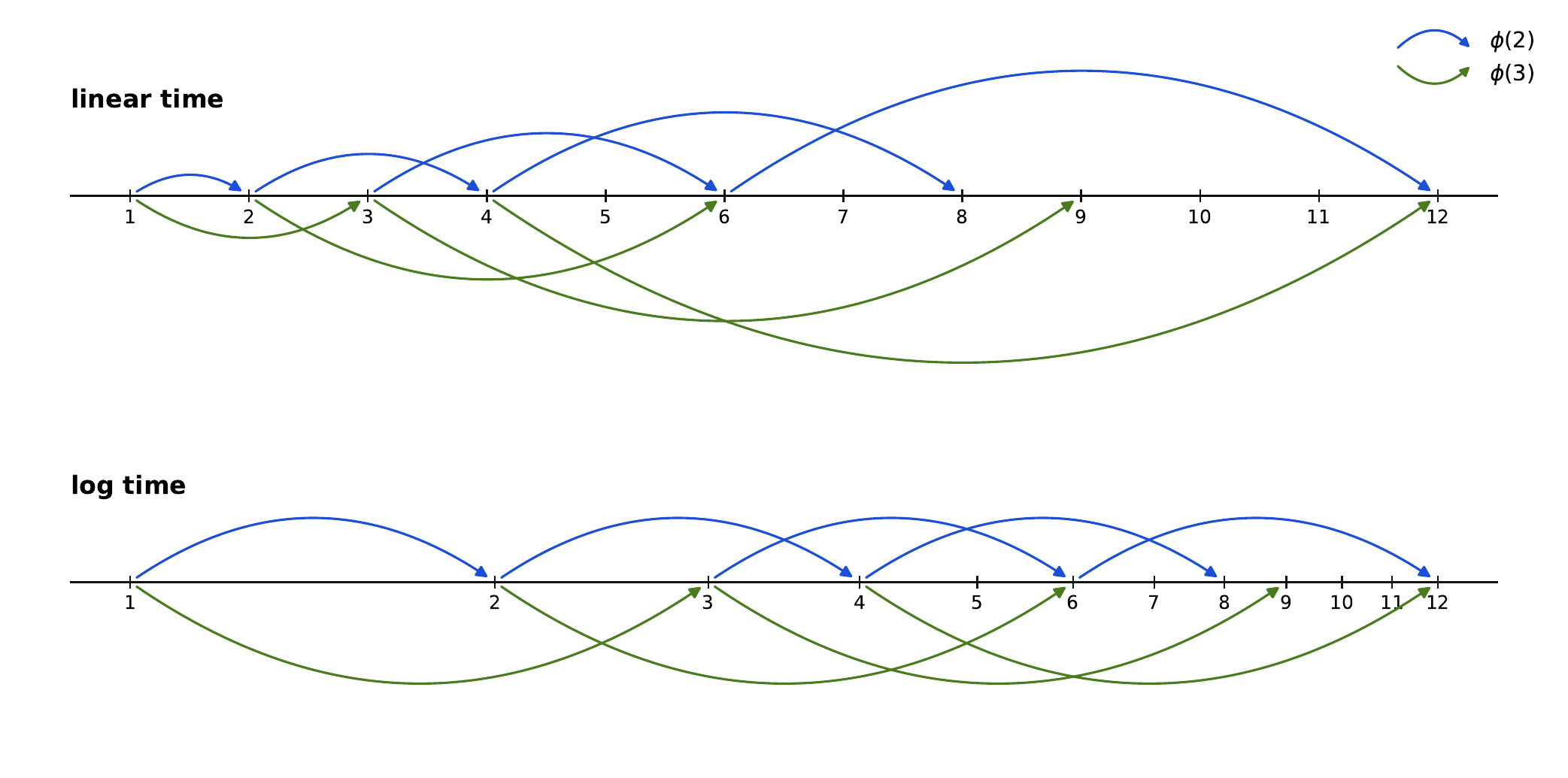}
  \caption{The logarithmic/exponential relation between tempo and harmonic rhythm in linear and log time.}
  \label{fig:linear-vs-log}
\end{figure}

In their discussion of aperiodicity, Hendriks et al.\ define periods in terms of equal differences of indices, which in our terminology is equal differences of time points. Consider the top panel of Figure~\ref{fig:linear-vs-log}, which spaces time points visually in this manner. While these sequences are aperiodic, there is substantial repetition as demonstrated by the numerous transpositions by $\phi(2)$ and $\phi(3)$, represented by blue and green arcs, respectively. More generally, measuring repeated harmonic motions of $\phi(\lambda)$ from
time point $q$ by linear time yields $\lambda q - q = (\lambda - 1)q$, where the linear time span of $T_{\phi(\lambda)}$ grows with $q$. In this case, while time points are fixed linearly, the harmonic rhythm decelerates logarithmically.

In the bottom panel, the situation is reversed by reckoning time in terms of ratios of time points, with equal ratios being equal steps of time. In log time, time points undergo an exponential acceleration, while harmonic motions are fixed linearly in time, since $\lambda q / q = \lambda$ is independent of $q$. Note also that the rationals are dense with respect to the integers, and we could draw an
infinitely dense number of arcs for any $\phi(\lambda)$. Thus, while this maximally self-similar structure is aperiodic in linear time, it is maximally periodic for all ratios $\lambda$ and time points $q$ in log time. Whether individual melodic realizations of this structure are periodic in log time depends solely on the selected time points and is independent of the freely chosen mapping of $\tau(p)$ on the primes.

\subsection{}
\label{sec:2.7}

Returning to prior work on self-similar melodies under single ratios, we can compare prior work with the current approach more clearly now. Amiot \cite{amiot} and Amiot, Agon, and Andreatta \cite{amiotagonandreatta} work with periodic melodies of period $n$, where taking every $\alpha$th note yields the same melody augmented by $\alpha$. The authors develop these ideas considerably beyond what can be surveyed here, and interested readers are encouraged to explore these papers further. What matters for our purposes is that their melodies repeat untransposed and that periodicity yields finite orbits, which is essential for their approach.

First, periodicity is essential for providing structure, because aperiodic melodies self-similar under a single ratio, $\alpha$, are nearly unconstrained. The only notes that are constrained are those whose indices are a multiple of $\alpha$. As an example, for a ratio of $3$, fully $2/3$ of the melody is unconstrained. (Compare to the current approach, in which only prime-numbered time points are unconstrained. Since we are working with rational time points, the density of unconstrained notes is zero, even for short melodies.) But by Hendriks et al.\ \cite{hendriks2012}, periodic sequences cannot give rise to maximal self-similarity, so the move to aperiodic lines in the current work is forced. Second, self-similarity in periodic melodies must be exact, untransposed copies of the original line. In maximally self-similar lines, untransposed copies force the trivial line of a constant pitch. Finally, as discussed in \S\ref{sec:2.6}, notions of periodicity are dependent on our measurement of time, with maximally self-similar structures being everywhere periodic in log time, while it is not clear what relevance this could have in Johnson's conception of self-similarity. As can be seen, Johnson's conception is not a
special case of the current approach---they are parallel developments.

% =====================================================================

\section{The Continuous Construction}
\label{sec:continuous}

\subsection{}
\label{sec:3.1}

The construction of \S\ref{sec:discrete} successfully generates melodic lines that are self similar at all positive rational time scales and thus can form canons at all rational tempo ratios while maintaining
harmonic consistency at coincident attacks. There are two problems remaining, however. First, this discrete approach does not allow for canons at irrational tempo ratios, which require extending the
construction to the continuous domain of the positive real numbers. Second, we have no harmonic control over notes that sound together without isochronous onsets. This includes notes with arbitrarily close attacks whose onsets may be perceptually (though not mathematically) simultaneous. Consider notes beginning at time points $1/4$ and $2/7$, which are separated by only $1/28$. (If we are progressing linearly through time points at one second per unit, this amounts to an inter-onset interval of a mere 36 milliseconds.) The directed interval between the pitches simplifies to $3\tau(2) - \tau(7)$. Since we are free to assign any value to $\tau(2)$ and $\tau(7)$ independently, there is no harmonic control over these near-simultaneous time points. While the \emph{free} mapping of $\tau(p)$ allows for an infinite variety of maximally self-similar structures, it comes at the expense of harmonic consistency beyond coincident attacks.

\subsection{}
\label{sec:3.2}

In order to extend our construction to irrational ratios and time points, we must redefine our homomorphism on the positive reals:
\begin{equation}
  \phi \colon (\mathbb{R}^{+}, \times) \longrightarrow (\mathbb{R}, +).
  \label{eq:hom-continuous}
\end{equation}
As discussed in the previous paragraph, the freely chosen mapping of the discrete approach leads to a problem of harmonic consistency, one that is exacerbated by extension to the irrationals. For an arbitrarily small time span, $\varepsilon$, we have a discontinuous function with a potentially infinite variety of intervals. In order to reclaim consistency, we require that this function be continuous, so that as the difference between two time points converges to zero, their mapping converges to the same pitch. By the logarithmic Cauchy functional equation, the only such solutions are
\begin{equation}
  \phi(q) \;=\; c \ln(q), \qquad q \in \mathbb{R}^{+}.
  \label{eq:cauchy}
\end{equation}
Of the infinite mappings in \S\ref{sec:discrete}, the only ones consistent with the condition of continuity are the family of mappings $\tau(p) = c \ln(p)$ that are proportional up to the scaling constant $c$.

\subsection{}
\label{sec:3.3}

As a concrete example consider the irrational time points $e$ and $\pi$. The directed interval is
\[
  \phi\!\left(\frac{\pi}{e}\right) = c \ln(\pi) - c \ln(e) = \phi(\pi) - \phi(e),
\]
and their respective pitches would be $T_{c \ln(e)}(s)$ and $T_{c \ln(\pi)}(s)$, for some reference pitch $s$.

Note, however, that in exchange for incorporating irrational ratios, we are accepting a greatly diminished variety of melodic lines, ones that increase monotonically and logarithmically. For example, taking $c = 12$, time points $q \in \mathbb{Z}^{+}$, a reference pitch of MIDI note 45 ($A_2$), and switching the base of the logarithm from $e$ ($\ln$) to $2$ ($\log_2$), our melodic line would be the harmonic series:
\[
  A_2,\; A_3,\; E_4\,(+2\text{\textcent)},\; A_4,\; C\sharp_5\,(-14\text{\textcent)},\; E_5\,(+2\text{\textcent}),\; \ldots
\]
Indeed, in our continuous construction any equal-spaced series of time points, $nx$ (where $n \in \mathbb{Z}^{+}$, $x \in \mathbb{R}^{+}$), amounts to a dilation of this one series.

\subsection{}
\label{sec:3.4}

\begin{figure}[htbp]
  \centering
  \includegraphics[width=0.5\textwidth]{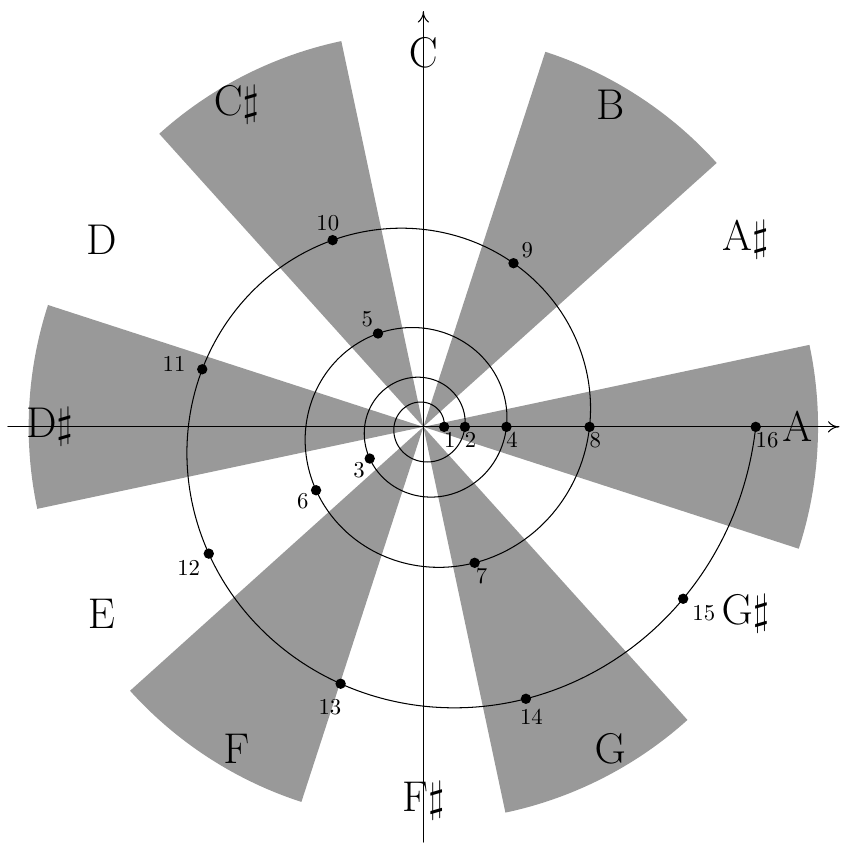}
  \caption{Harmonic series plotted on the logarithmic spiral $r = 2^{\theta/2\pi}$ and approximated by the chromatic scale. Octave placement for pitches is given by the winding number.}
  \label{fig:spiral}
\end{figure}

In order to combine the benefits of the continuous function with the flexibility of the discrete approach, we will quantize the continuous output and map the discretized result onto a repeating harmonic cycle.
Figure~\ref{fig:spiral} demonstrates the process using the familiar harmonic series. First, we transform the logarithmic function into the logarithmic spiral $r = 2^{\theta/2\pi}$, so that the spiral doubles
with every complete turn. (This constant rate of expansion and resulting self-similar structure is a defining feature of all logarithmic spirals, first described by Descartes in 1638 and extensively studied by Jacob Bernoulli, who called the curve the \emph{spira mirabilis}~\cite{archibald1920}.) The points on this spiral are placed at integer values for $r$, with the radius being interpreted as time points. (We will consistently switch from $q$ to $r$ when discussing time points from here on. While the figure is restricted to integer time points, $r$ is free to be any positive real number.) The circle is divided into 12 regions, each mapping to a different pitch class, beginning with $A$ on the right and ascending chromatically moving counter-clockwise. Each time point is quantized to the region in which it lies and is subsequently mapped to the region's associated pitch class. Octave placement (and thus pitch) can be deduced from the winding number of the spiral. From the figure, we can construct a harmonic series as it is typically approximated in equal temperament. Assuming we have $A_2$ as the pitch of time point 1, we have the same sequence of pitches as \S\ref{sec:3.3} without microtonal deviations. Note that the arc lengths between successive time points remain constant while their
angular displacement decreases logarithmically. This is the geometric representation of the logarithmic relation between time points and harmonic rhythm discussed in \S\ref{sec:2.6}.

\subsection{}
\label{sec:3.5}

In this subsection we adapt the homomorphism $\phi(r) = c \log_\lambda(r)$ to incorporate the quantization process outlined in \S\ref{sec:3.4} as a new function of directed interval, $\tilde{\phi}$. As a caution before proceeding, while in Figure~\ref{fig:spiral} the regions correspond to steps of the chromatic scale and a complete cycle to an octave, in general this is not the case. What is necessary in order to maintain maximal self-similarity is that the regions correspond to a generating interval, $g$. In the case of the ascending chromatic scale or cycle of equal-tempered perfect fifths, we have $g = 1$ and $g = 7$, respectively. We could also have $g = 12 \log_2(3/2)$ for pure fifths, where a cycle of 12 fifths differs from seven octaves by the Pythagorean comma, or $g = 1, 2, 3$, etc., corresponding to the generic intervals of a diatonic or other scale.

Another caution concerns the emphasis on pitch rather than pitch class in \S\ref{sec:discrete}. In the discrete construction, we can freely choose a balance of ascending and descending intervals to ensure that the resulting melodic line remains within a usable musical range. We are not obligated to work with pitch for the purpose of maintaining harmonic consistency, since octave-reduced intervals, e.g.\ octaves and unisons, tenths and thirds, are generally regarded as equivalent. In the continuous construction, however, we are forced to use a monotonic function, and strictly preserving octave information would yield uninteresting monotonic lines. For this reason, we will generally work with pitch classes and pitch-class intervals below.

We begin by rounding $\phi(r)$ to the nearest integer, which yields the appropriate region for each time point, and multiply this region index by the generating interval: $g\,[\,c \log_\lambda r\,]$. However, the regions may be offset by a given amount, $\psi$, and not precisely centered with respect to the axes, as can be seen in Figure~\ref{fig:spiral}.\footnote{The reason for this slight offset in Figure~\ref{fig:spiral} is to accommodate alternate spellings of the thirteenth harmonic. While the thirteenth harmonic of an $A$ fundamental lies closest to $F\natural$, it is often approximated by $F\sharp$ in sonorities and scales relating to the harmonic series, such as Scriabin's Mystic chord and the acoustic scale. With the offset in Figure~\ref{fig:spiral}, time point 13 sits in the $F\sharp$ region; without the offset, it would lie inside the $F\natural$ region. An analogous offset plays a related role in the circle-of-fifths example discussed in \S\ref{sec:examples}.} Pulling all this together, the function for directed intervals, taking into account the quantization process, is
\begin{equation}
  \tilde{\phi}(r) \;=\; g\,\bigl[\, c \log_\lambda r - \psi \,\bigr].
  \label{eq:phi-tilde}
\end{equation}
If using pitch classes instead of pitches, we take $\tilde{\phi}(r)$ modulo the size of the octave.\footnote{In Figures~\ref{fig:spiral} and~\ref{fig:fifths} we identify each region by a pitch class, which is a convenient abuse of notation, since regions are properly directed intervals. This shorthand is only possible in these figures because the harmonic cycles are commensurate with an integral number of octaves.}

% =====================================================================

\section{Circle-of-Fifths examples}
\label{sec:examples}

\begin{figure}[htbp]
  \centering
  \includegraphics[width=\textwidth]{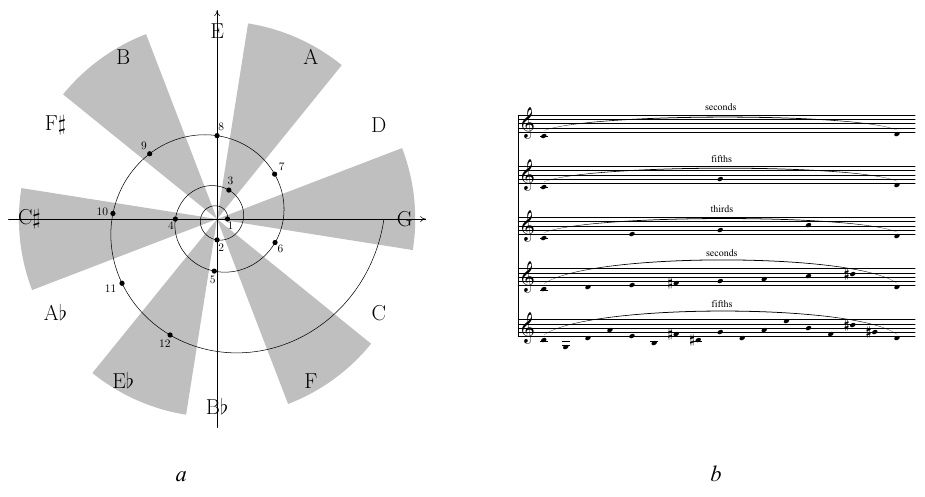}
  \caption{(a) Time points represented by a logarithmic spiral against a background harmonic cycle of perfect fifths. Time points six through eleven yield the sequence $C$--$D$--$E$--$B$--$C\sharp$--$A\flat$. (b) Progressive embellishment of the intervals associated with time points six through eleven.}
  \label{fig:fifths}
\end{figure}

The following examples are based on a single derived melody, with a background harmonic cycle of an ascending circle of fifths. (See Figure \ref{fig:fifths}.) Superimposed on this cycle is a logarithmic spiral $r = 2^{\theta / (3\pi/2)}$, in which $r$ doubles with every three quarters of a turn rather than with every complete turn. Taking the integer values of $r$ between six and eleven, the corresponding notes are $C$--$D$--$E$--$B$--$C\sharp$--$A\flat$, which form the beginning of the derived melody. For the remainder of this melody, the intervals are embellished in a more typically musical manner, with seconds being divided into fifths, fifths into thirds, thirds into seconds, and seconds again being divided into fifths.\footnote{For audio and more examples, see~\cite{infinitecanons}.}

It is worth noting the significant positive offset in the regions of Figure~\ref{fig:fifths}(a). Without this offset, time point 9 would lie in the $F\sharp$ region, yielding the sequence $C$--$D$--$E$--$F\sharp$--$C\sharp$--$A\flat/G\sharp$, which switches abruptly from major seconds to fifths. With the offset, the line more gradually transitions from seconds to fifths, better reflects the smooth underlying logarithmic sampling of the circle of fifths, and is more musically compelling in a way that propagates through the subsequent embellishments, gradually transitioning from fifths to thirds to seconds and back to fifths.

\begin{figure}[htbp]
  \centering
  \includegraphics[width=\textwidth]{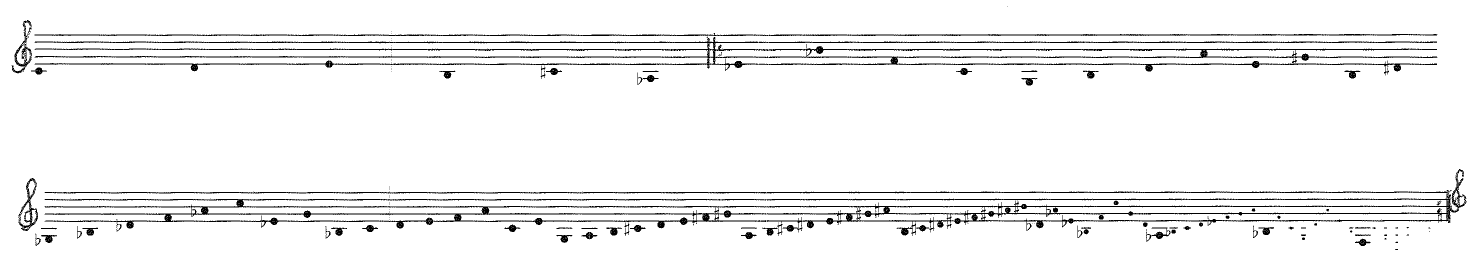}
  \caption{Canon line resulting from time points six through twelve and the diminution process shown in Figure~\ref{fig:fifths}.}
  \label{fig:canon-line}
\end{figure}

The resulting melody is shown in Figure \ref{fig:canon-line}. Over the duration of each half line, the tempo increases by a factor of two, so that while there are six notes in the first half of the first line, there are 12 in the second half, 24 in the first half of the second line, and 48 in the second half of the second line. (The absolute length of each note on the page corresponds to its absolute sounding duration.) This is a very salient difference in presentation of the self similar structure. In the example of \S\ref{sec:discrete}, the tempo remained constant while the repeating directed intervals varied with respect to time. In the examples of this section, it is the background that moves at a fixed rate, while the tempo varies. Though the underlying structure is the same, the melodies seem to have very little in common.\footnote{For more on Nancarrow's own approach to acceleration canons and a more generalized approach to continuous variable tempos, see Callender~\cite{callenderPNM}.}

By the midpoint of the second line, the melody is moving eight times as fast as the beginning. Over the last quarter of the melodic line, every eighth note remains at a constant volume while all other notes fade out (represented by decreasing note head size), gradually revealing the opening of the melody moving at the same tempo transposed down a minor third. The portion of the line between the repeat signs can thus repeat \emph{ad infinitum}, down a minor third with each repetition.

\begin{figure}[htbp]
  \centering
  \includegraphics[page=1, width=\textwidth]{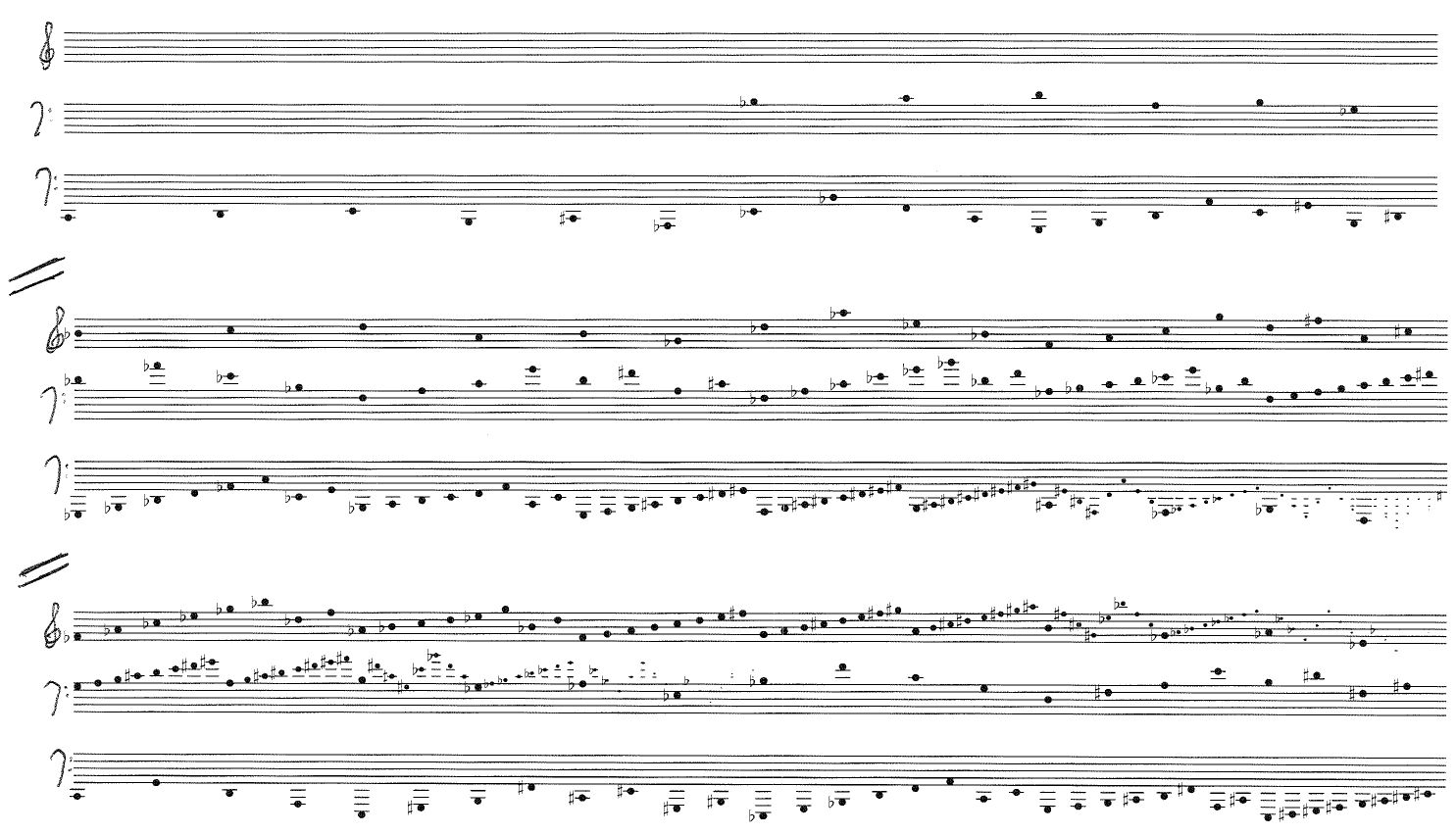}
  \includegraphics[page=2, width=\textwidth]{canon-4-2-1.pdf}
  \caption{Canon $4\!:\!2\!:\!1$.}
  \label{fig:canon421}
\end{figure}

We begin with a relatively straightforward three-voice canon in the ratios $4\!:\!2\!:\!1$ in Figure \ref{fig:canon421}, where notes starting simultaneously always form a root-position triad. In this case the fastest voice begins alone. Since the canon line is accelerating, the second voice enters once the first voice has accelerated by a factor of two, which occurs halfway through the first line of the melody above. Similarly, the third voice will enter when the first voice has accelerated by a factor of four, which occurs at the end of the first line of the melody. Since the acceleration of each voice is exponential, the time delay between voices remains constant even though the voices are in different tempos.

\begin{figure}[htbp]
  \centering
  \includegraphics[page=1, width=\textwidth]{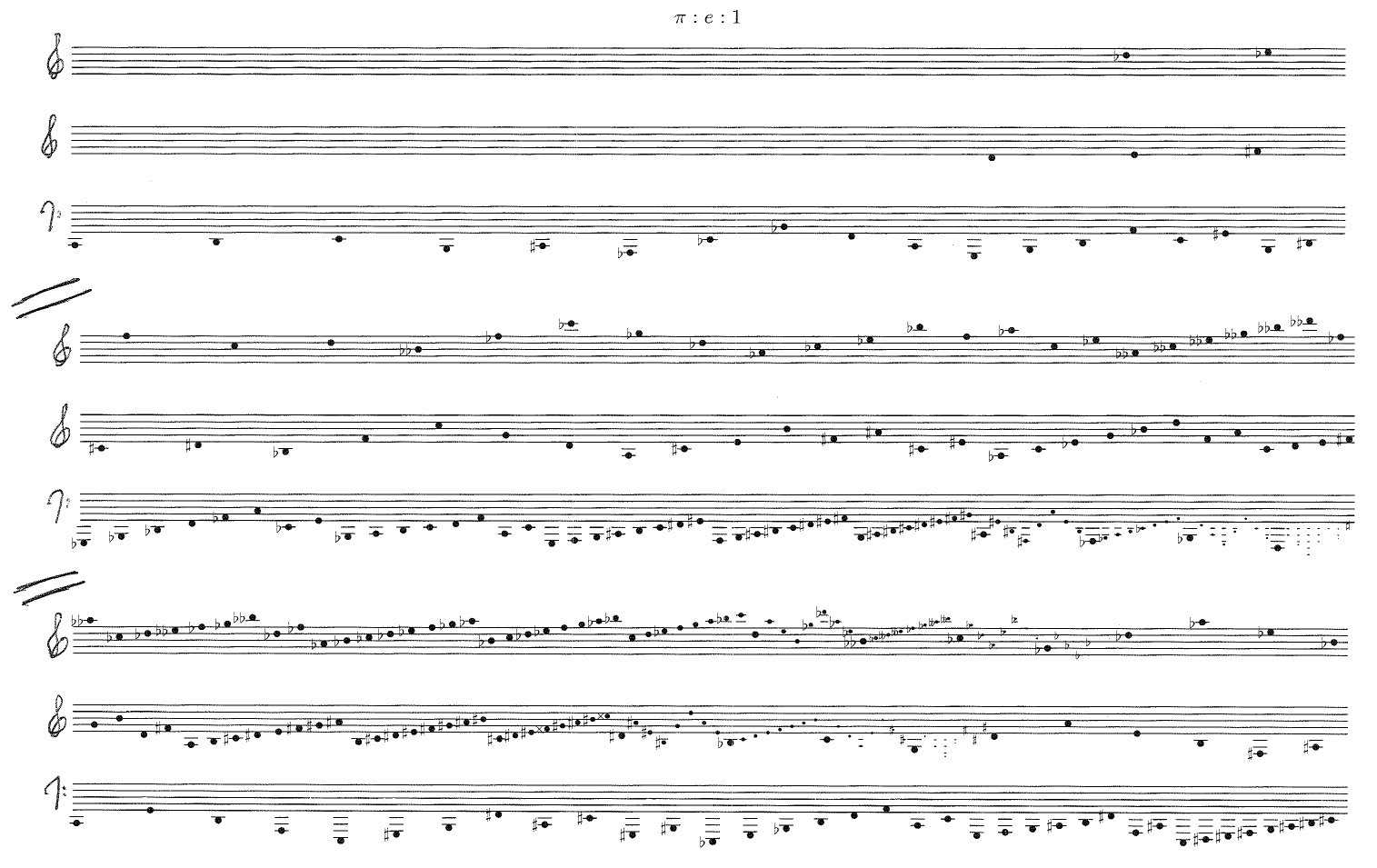}
  \includegraphics[page=2, width=\textwidth]{canon-pi-e-1.pdf}
  \caption{Canon $\pi\!:\!e\!:\!1$.}
  \label{fig:canon-pie1}
\end{figure}

Due to their continuous construction, these canons work just as well with irrational tempo ratios. Figure \ref{fig:canon-pie1} is a three-voice canon in the ratios $\pi\!:\!e\!:\!1$, in which theoretically no notes will begin simultaneously. (Practically speaking, some notes will begin close enough to be perceived as simultaneous.) In this case, the predominant chord is again the root-position triad. What this means is that the closer two notes are to beginning simultaneously, the more likely they are to form part of a triad. Simultaneously sounding notes in which the attacks are farther apart may form various kinds of incomplete diatonic seventh and ninth chords. Since each voice is progressively elaborating the same cycle of ascending fifths in a diatonic manner, the voices will always ``agree'' as to the prevailing diatonic collection, even as this collection smoothly passes through all of its twelve transpositions.\footnote{For precedents of irrational tempo canons, see Nancarrow's Studies Nos. 33, 40, and 41. See also Paul Usher's arrangement of Study No. 33, $\sqrt{2}/2$, for the Arditti Quartet, discussed in Callender~\cite{callenderMTO}. For more on Nancarrow's tempo canons, their structural properties, and tempo dissonance see Gann~\cite{gann}, Nemire~\cite{nemire}, and Thomas~\cite{thomas2000pnm}~\cite{thomas}.}

% =====================================================================

\section{Table Canons}
\label{sec:table}

An initially surprising consequence of maximally self-similar lines is that they naturally give rise to canons by retrograde inversion, also known as table canons, where the line is combined with itself backward and upside down.\footnote{The score for such a canon can be placed on a table and read by two performers on opposite sides, giving rise to the term \emph{table canon}. One example is often attributed to Mozart, \emph{Der Spiegel} (The Mirror) for two violins.} An intuitive understanding of this potential is provided by the geometric properties of the logarithmic spiral, though table canons arise in both the discrete and continuous constructions, as we will shortly demonstrate algebraically.

\begin{figure}[htbp]
  \centering
  \includegraphics[width=0.4\textwidth]{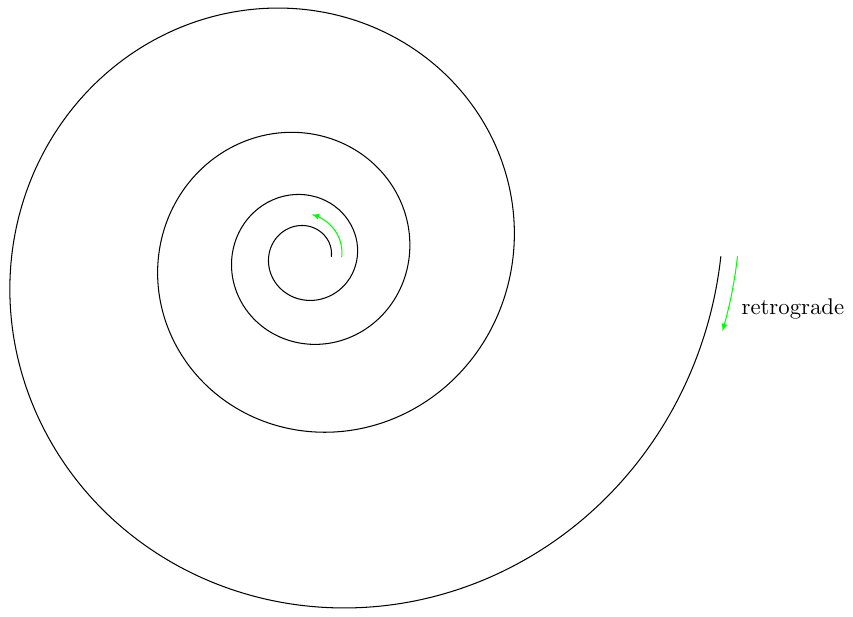}
  \caption{The retrograde of the original theme moves in the opposite direction through the harmonic cycle.}
  \label{fig:retrograde-spiral}
\end{figure}

Consider Figure~\ref{fig:retrograde-spiral}. If the original melody moves along the logarithmic spiral from the center out in the counterclockwise direction, then the retrograde of the melody will move from the outside toward the center in the clockwise direction. The original melody cannot be combined with its retrograde because the two are moving in opposite directions through the harmonic cycle; e.g., if the original moves through the ascending circle of fifths, then its retrograde will move through the \emph{descending} circle of fifths, and we will be unable to maintain harmonic consistency.

\begin{figure}[htbp]
  \centering
  \includegraphics[width=0.4\textwidth]{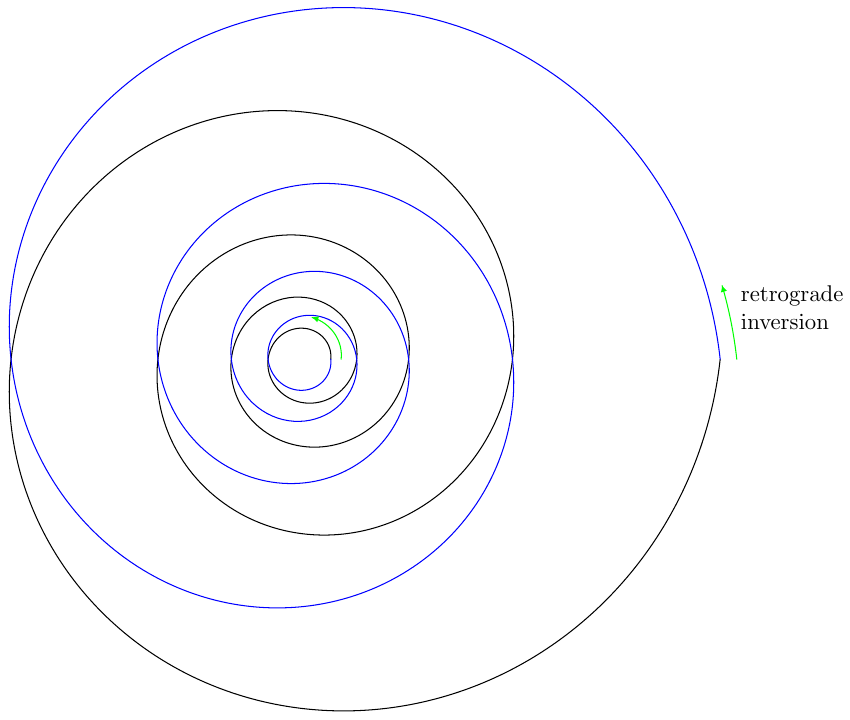}
  \caption{The retrograde inversion of the original theme moves in the same direction through the harmonic cycle.}
  \label{fig:ri-spiral}
\end{figure}

Similarly, the \emph{inversion} of the original melody will proceed in the opposite direction. We can represent the inversion by inverting the spiral about the $x$-axis, shown in Figure~\ref{fig:ri-spiral} by the blue spiral, where the inversion of the melody would move from the center out in the clockwise direction. However, as shown in the figure, the \emph{retrograde inversion} moves in the counterclockwise direction and thus moves in the same direction through the harmonic cycle as the original melody; e.g., if the melody moves through an ascending circle of fifths, then its retrograde inversion will also move through the ascending circle of fifths.

We proceed with a couple of examples featuring the melody from \S\ref{sec:examples}, now in canon by retrograde inversion. In the first example, the two forms of the melody begin and end together in a two-voice table canon. However, the original version continually accelerates, while the retrograde inversion continually decelerates, preserving the harmonic rhythm at a fixed tempo (as discussed in \S\ref{sec:2.6}). Throughout, the predominant interval is the tenth.

\begin{figure}[htbp]
  \centering
  \includegraphics[page=1, width=\textwidth]{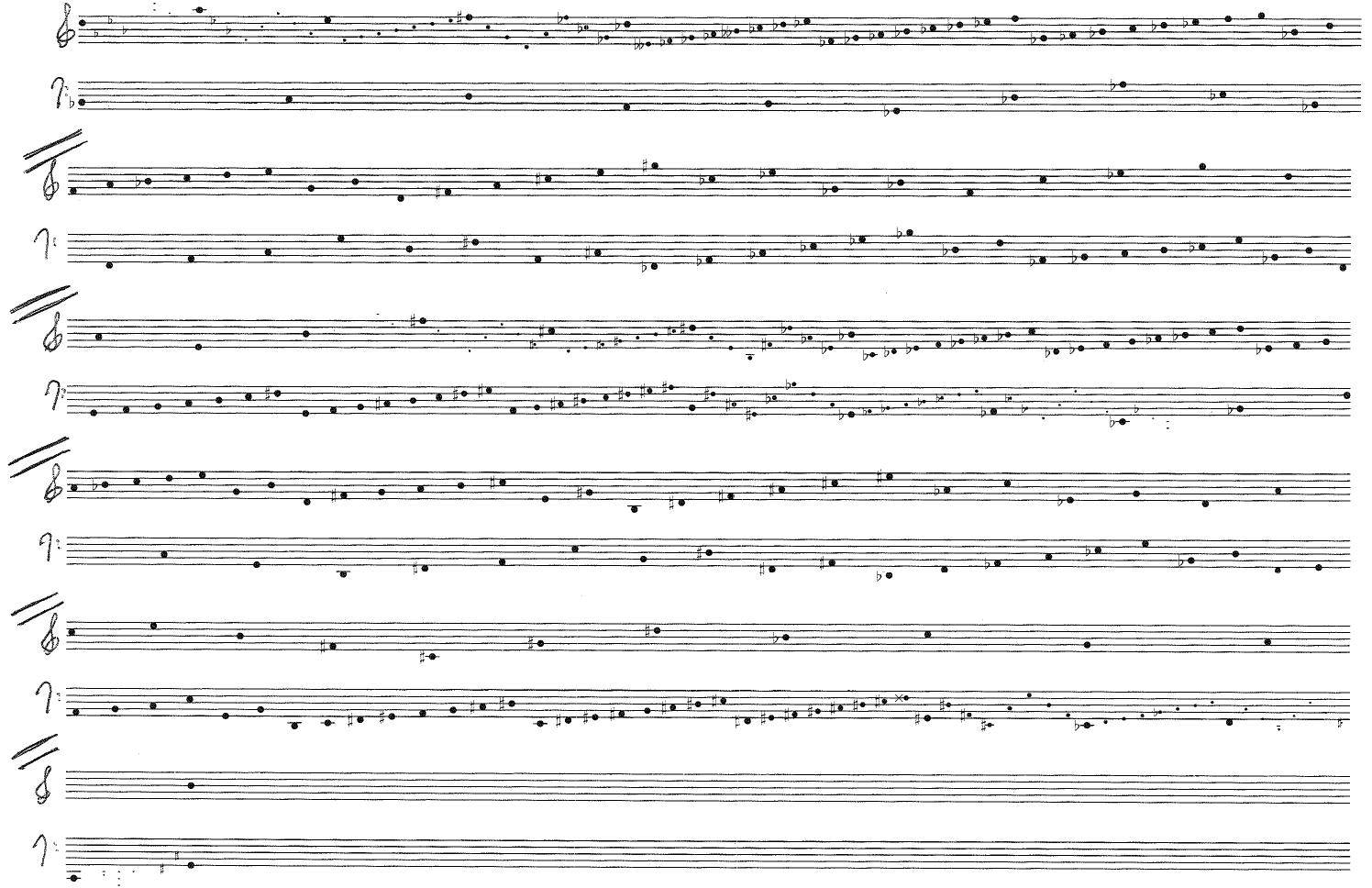}
  \caption{Table canon (canon by retrograde inversion).}
  \label{fig:table-canon}
\end{figure}

Our final canon is a six-voice table canon which combines a three-voice canon in the ratios $28\!:\!24\!:\!21$ with its three-voice retrograde inversion. Within each of the three-voice canons the predominant interval is the octave, while \emph{between} the retrograde inversions the primary interval is the tenth. Thus this canon sounds like a texturally enriched version of the preceding canon.

\begin{figure}[htbp]
  \centering
  \includegraphics[page=1, width=4.5in]{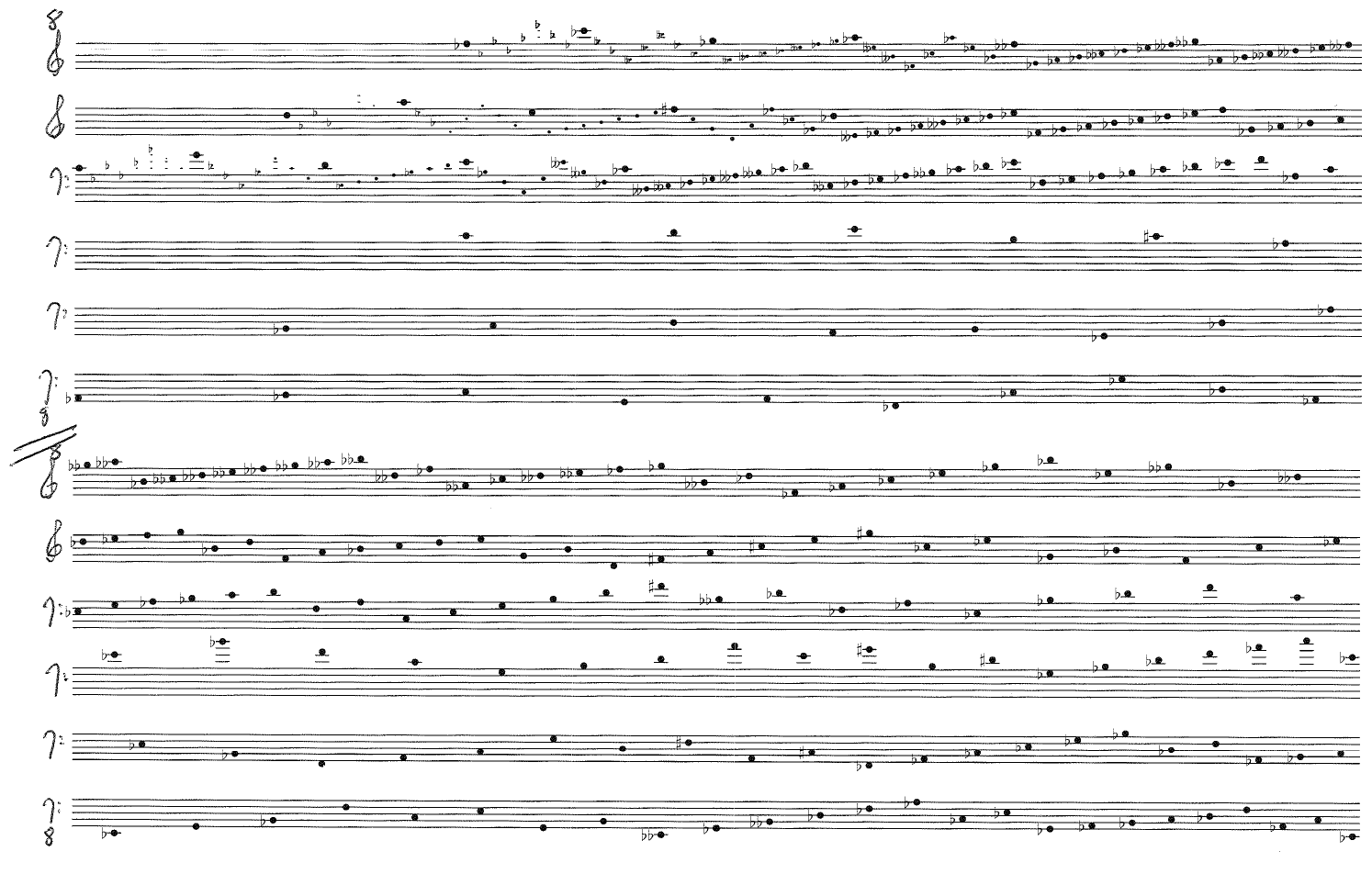}
  \includegraphics[page=2, width=4.5in]{table-canon-28-24-21.pdf}
  \includegraphics[page=3, width=4.5in]{table-canon-28-24-21.pdf}
  \caption{Table canon $28\!:\!24\!:\!21$.}
  \label{fig:table-canon-28}
\end{figure}

Algebraically, this self-similarity by retrograde inversion is a direct consequence of our homomorphism for directed intervals, $\phi$. Recall the discussion in \S\ref{sec:2.6} of the difference between linear and log time, with the difference between two time points defined by their ratio in the latter. A retrograde in log time, what one might call a \emph{proportional retrograde}, maps time point $r$ to time point $t_0^{2}/r$, where $t_0$ is the fixed temporal axis of the retrograde. For example, if the axis is $t_0 = 4$ and we have the sequence $r = 1, 2, 3, 4, 5, 6, 7, 8$, then the proportional retrograde is $\frac{t_0^{2}}{r} = 16,\; 8,\; \tfrac{16}{3},\; 4,\; \tfrac{16}{5},\; \tfrac{8}{3},\; \tfrac{16}{7},\; 2 .$ The directed intervals for the retrograde are
\begin{equation}
  \phi\!\left(\frac{t_0^{2}}{r}\right) = 2\phi(t_0) - \phi(r).
  \label{eq:proportional-retrograde}
\end{equation}
In other words, the proportional retrograde yields a transposed ($2\phi(t_0)$) inversion ($-\phi(r)$) of the original line. (Note also how this equation relates to musical inversion of a pitch, $p$, about an axis, $n$: $2n - p$.) Reflection in time corresponds to a reflection in pitch, and as a consequence, these lines are inversionally symmetric (up to transposition and in log time) about every time point.

If we take the inversion of the retrograde, we have
\begin{equation}
  -\phi\!\left(\frac{t_0^{2}}{r}\right) = \phi(r) - 2\phi(t_0),
  \label{eq:ri-collapse}
\end{equation}
which is simply a transposed version of the original---under the conditions of maximal self-similarity, retrograde inversion collapses to transposition. As demonstrated by the above examples, melodic realizations of this structure remain perceptually distinct under retrograde inversion, provided the line is not based solely on a logarithmic series of time points. It is the underlying structural equivalence that makes the complex combinations of these lines possible.

% =====================================================================

\section{Conclusion}
\label{sec:conclusion}

We have constructed maximally self-similar lines in (1) the discrete case, where positive rational time points are mapped to real-valued directed intervals through prime factorization and a freely chosen map from the primes to the reals; and (2) the continuous case, where real-valued time points are associated with directed intervals through quantized logarithms mapped to harmonic cycles. The main result in both cases is that the vertical interval between voices in a tempo ratio of $\lambda_i / \lambda_j$ is given by the homomorphism $\phi(\lambda_i / \lambda_j)$, which is a constant independent of time. Furthermore, we demonstrated that under these constructions retrograde inversion collapses to transposition. The musical payoff of the mathematics is the ability to compose melodic lines that combine in any number of voices, in any tempo ratios, either forward or in retrograde inversion while maintaining harmonic consistency---lines with an infinite number of canonic combinations.

Future work will proceed along a number of lines. First, there are many avenues to explore other musical realizations (including a variety of harmonic cycles),\footnote{These realizations include \emph{Canonic Offerings}, for string quartet, invited by Dmitri Tymoczko for a concert as part of the 2013 Bridges Conference: Mathematics, Music, Art, Architecture, Culture in Enschede, The Netherlands and \emph{spira mirabilis: infinite tempo canons}, for player piano, invited by Benjamin Broening for 2018 Third Practice Music Festival at the University of Richmond, VA.} multimedia presentations, an interactive app, and relation to recent work by Tymoczko~\cite{tymoczko2026}. Second, it is worth considering the relationships between the current approach and that of Johnson, Amiot, and others in greater detail than \S\ref{sec:2.7}.  Third, the pitch symmetry about every time point noted toward the end of \S\ref{sec:table} suggests an alternative formalization in terms of reflection. Fourth, it would be helpful and musically suggestive to consider the move from the discrete to the continuous case in greater detail: one can incorporate irrational ratios within the construction of \S\ref{sec:discrete} through Hamel bases at the expense of potentially wildly discontinuous structures, creating problems for the harmonic consistency that motivated the requirement for continuity in \S\ref{sec:3.2}, while the quantization process may also present difficulties for maintaining harmonic control. Relatedly, it is important to clearly identify the effects of rounding errors alluded to at the end of \S\ref{sec:examples}. Fifth, the function $\phi$ invites discussion of homomorphisms between Lewinian generalized interval systems~\cite{lewin1987} (following Tymoczko's response~\cite{tymoczko2008} to Hook~\cite{hook2007}), particularly the GISes corresponding to products over the positive rationals and addition over the reals discussed in Lewin~\cite{lewin1987}. Finally, proof of the expanded aperiodicity results in \S\ref{sec:2.5} is forthcoming.

\end{document}